\documentclass[twocolumn,showpacs,preprintnumbers,amsmath,amssymb]{revtex4}
\usepackage{dcolumn}
\usepackage{bm}
\expandafter\ifx\csname package@font\endcsname\relax\else
 \expandafter\expandafter
 \expandafter\usepackage
 \expandafter\expandafter
 \expandafter{\csname package@font\endcsname}%
\fi

\usepackage{graphicx}   
\usepackage{amsmath}    

\newcommand{\bra}[1]{\mbox{$\langle #1 |$}}
\newcommand{\ket}[1]{\mbox{$| #1 \rangle$}}

\begin{document}

\title{Nonorthogonal decoy-state Quantum Key Distribution }

\author{Jing-Bo Li, and Xi-Ming Fang\footnote{corresponding author: Email: fxm601@yahoo.com.cn}}

\affiliation{%
Department of Physics, Hunan Normal University, Changsha 410081, People's Republic of China\\
}%

\date{\today}

\begin{abstract}
In practical quantum key distribution (QKD), weak coherent states
as the photon source have a limit in secure key rate and
transmission distance because of the existence of multiphoton
pulses and heavy loss in transmission line. The decoy-state method
and the nonorthogonal encoding protocol are two important weapons
to combat these effects. Here, we combine these two methods and
propose an efficient method that can substantially improve the
performance of QKD. We find a 78 km increase over the prior record
using the decoy-state method and a 123 km increase over the result
of the SARG04 protocol in transmission distance.
\end{abstract}

\maketitle

Quantum key distribution (QKD)\cite{gisin,bene} allows two users,
Alice and Bob, to communicate in absolute security in the presence
of an eavesdropper, Eve. Unlike conventional cryptography, the
security of QKD is based on the uncertainty principle and the
noncloning theorem \cite{wz}. In other words, the measurement of
an unknown quantum state modifies the state itself. Thus, Eve
cannot gain any information on the key without introducing any
error in the correlations between Alice and Bob. However, in
practical implementations, an attenuated laser pulse (a weak
coherent state)is often used as the source. The existence of
multiple photon pulses, even though very rare, poses a serious
problem for the security of the protocol, especially in high lossy
channel. An eavesdropper (Eve) can in principle have the full
information of Bob's sifted key by using the
photon-number-splitting (PNS)attack \cite{higm,luken,lukenetc}:
Eve blocks all single-photon pulses and part of multi-photon
pulses and separates each of the remained multi-photon pulses into
two parts therefore each part contains at least one photon. She
keeps one part and sends the other part to Bob, through a lossless
channel.

Recently, two important methods have been proposed to overcome PNS
attacks. One is the decoy-state method firstly proposed by Hwang
\cite{Hwang}, and further studied by Wang \cite{Wang}, and also Lo
and co-workers \cite{lmc,mqzl}. Particularly, by combining the
idea of the entanglement distillation approach by Gottesman, Lo,
Lutkenhaus, and Preskill (GLLP) \cite{GLLP} with the decoy state
method, they achieved a formula for secure key generation rate
\cite{lmc}:
\begin{equation} \label{practicalkeyrate}
     S \geq q  \{-Q_{\mu}f(E_{\mu})H_2(E_{\mu})+Q_1[1-H_2(e_1)]\},
\end{equation}
where $q$ is the sifting efficiency depending on the
implementation (1/2 for the BB84 protocol, because half the time
Alice and Bob bases are not compatible), $Q_{\mu}$ and $E_{\mu}$
are the gain (i.e., counting rate \cite{Wang}) and quantum bit
error rate (QBER) of the signal state respectively, and can be
measured directly, $Q_1$ and $e_1$ are the gain and QBER of
single-photon states respectively, and can be estimated by using
decoy state method, $f(E_{\mu})$ is the error correction
efficiency \cite{cascade}, and $H_2$ is the binary Shannon
entropy, given by:
\begin{equation} \label{entropy}
H_2(x) =  -x\log_2(x)-(1-x)\log_2(1-x).
\end{equation}
The other is the nonorthogonal states encoding protocol proposed
by Scarani, Acin, Ribordy and Gisin (SARG04) \cite{SARG04}, which
uses exactly the same four states as in BB84 \cite{bene}, and only
the classical sifting procedure is different from BB84: instead of
revealing the basis, Alice announces publicly a pair of
nonorthogonal states. Thus, Eve needs at least three photons to
obtain full information. This means one can utilize the two-photon
part to generate a secure key. However, either the decoy state
method or the nonorthogonal states encode protocol has no further
security analysis on it.

In this paper, we first present a simple method that can study the
secure key generation rate when single-photon and two-photon
pulses are employed to generate secure key. The structure of the
paper is as follows. First, we derive a formula for secure key
generation rate, where two-photon part is included. Next we
present a simple method that will give a tight bound to $Q_0$,
$Q_1$, $e_1$, $Q_2$ (the gain of two-photon states)and $e_2$ (the
error rate of two-photon states) respectively. Then we present the
advantage of this new protocol at secure key generation rate and
transmission distance by comparing with the results in \cite{lmc}.
Finally, we discuss and conclude.

{\em Our new GLLP formula.} The secure generation rate must
include the two-photon part when we use SARG04 protocol. So we
need to modify Eq. \eqref{practicalkeyrate} to satisfy our
purpose.

{\bf Theorem} The key generation of an nonorthogonal encoding
scheme is given by:
\begin{equation}\label{newkeyrate}
 S \geq q  \{-Q_{\mu}H_2(E_{\mu})+Q_0+Q_1[ 1- H_2(e_1)]+Q_2[ 1-
H_2(e_2)]\},
\end{equation}
where $q$ is 1/4 for SARG04, and $Q_0$ is the gain of the vacuum
signals.

Now, let us prove it. According to the Csisz\'ar-K\"orner theorem
\cite{cktheorem}: if the mutual information Alice-Bob is larger
than either the mutual information Alice-Eve or Bob-Eve, then
Alice and Bob can distil a secret key. The secure key generation
in QKD satisfies
\begin{equation}\label{s}
  S \geq I(A:B)-I(B:E),
\end{equation}
where $I(A:B)$ and $I(B:E)$ are  mutual information of Alice-Bob
and Bob-Eve respectively, and are given by:
\begin{eqnarray}\label{mutualinf}
   I(A:B) &=& qQ_{\mu}(1-H_2(E_{\mu}))\\
   I(B:E) &=& q\{Q_1H_2(e_1) + Q_2H_2(e_2) + \sum_{n\geq 3}Q_n\},
\end{eqnarray}
where $Q_n$ is the gain of n-photon states, and q is 1/4 for
SARG04. The vacuum signals do not contribute to it at all because
of the mutual information of vacuum being zero. Here, we take the
most conservative assumption that Eve has all the information on
all tagged pulses (the parts for photon number $n \geq 3$) and
obtains full information stemming from the QBERs $e_1$ and $e_2$.
Combining Eq. \eqref{s} and $Q_{\mu}=\sum_{n\geq 0}Q_n$, we get
the result of our theorem. In fact Eq. \eqref{newkeyrate} can be
generalized from Lo's theorem \cite{lovacumm} directly if only
adding the secure generation rate of the two-photon part. As
discussed in \cite{lmc}, practical error correction protocols are
generally inefficient. Thus, the secure key generation rate for
practical protocols is given by:
\begin{eqnarray}\label{newrealkeyrate}
 S \geq q  \{-Q_{\mu}f(E_{\mu})H_2(E_{\mu})+Q_0+Q_1[ 1- H_2(e_1)]\nonumber \\
 +Q_2[ 1-H_2(e_2)]\}.
\end{eqnarray}

{\em The optimal secure key generation rate without decoy states.}
Although we have obtained the Eq. \eqref{newrealkeyrate} that can
calculate the  secure key generation rate for the SARG04 protocol,
we have to discard the it due to the presence of Eve. In this
case, Eve can block all single-photon pulses or all two-photon
pulses, she can get more information, so the worst secure key
generation rate is given by
\begin{equation}\label{worst}
 S_{worst} = \frac{1}{4}(-Q_{\mu}f(E_{\mu})H_2(E_{\mu})+Q_0+ \Omega Q_{\mu}[ 1-
 H_2(\frac{e}{\Omega})]),
\end{equation}
where $f(E_{\mu})=1$ for convenience, and $\Omega$, the fraction
of untagged photons, satisfies
\begin{equation}\label{upper}
 \Omega=1- \frac{(1+\mu+\mu^2/2)e^{-\mu}}{Q_{\mu}}.
\end{equation}
$S_{worst}$ is optimised if we choose $\mu=\mu_{optimal}$, which
fulfills
\begin{eqnarray}\label{optimalmu}
 \eta e^{-\eta \mu_{optimal}} = \frac{1}{2}\mu_{optimal}^2
e^{-\mu_{optimal}}.
\end{eqnarray}
Since for realistic setup we expect that $\eta\ll 1$, we find
$\mu_{optimal} \approx \sqrt{2\eta}$.

{\em The lower bound of the secure key generation rate with decoy
states.} A verified lower bound of secure key generation rate can
be obtained by using decoy-state method. This method is dependent
on the real-world QKD protocols deeply. In practical
implementations, a weak coherent state (i.e., a dephased coherent
state) is a mixed state of
\begin{equation}\label{density}
\rho=\int\frac{d\theta}{2\pi}\ket{\sqrt{\mu}
e^{i\theta}}\bra{\sqrt{\mu} e^{i\theta}}=\sum_nP_n(\mu)\ket{\mu}
\bra{\mu},
\end{equation}
where $P_n(\mu)=\frac{\mu^ne^{-\mu}}{n!}$ and $\mu$ is the mean
photon number. The gain, $Q_{\mu}$, and QBER, $E_{\mu}$, are given
by
\begin{eqnarray} \label{Decoy:QSingal}
Q_{\mu} &=& \sum_{n \geq 0}Q_n\\
Q_{\mu}E_{\mu} &=&  \sum_{n \geq 0}Q_ne_n
\end{eqnarray}
and $Q_n=Y_nP_n(\mu)$, where $e_n$ and $Y_n$ are respectively the
error rate and yield of the n-photon state. In the normal case
that there is no eavesdropper, $Q_{\mu}$ and $E_{\mu}$ are given
by \cite{mqzl}:
\begin{eqnarray} \label{normalqe}
Q_{\mu} &=& Y_0 + 1 - e^{-\eta\mu},\\
Q_{\mu}E_{\mu} &=&  e_0Y_0 + e_{det}(1-e^{-\eta\mu}),
\end{eqnarray}
where $e_{det}$ is the probability that a photon hit the erroneous
detector, $\eta$ is the overall transmission probability of a
photon.

In the presence of an eavesdropper, Eve, we can use the
decoy-state method to detect Eve's attacks. The essence of decoy
state idea is that Eve cannot distinguish the decoy state from the
signal state. So the signal state and the decoy state have the
same values for the yield, $Y_n$, and QBER, $e_n$. In order to
achieve the unconditional security of QKD with the key generation
rate given by Eq. \eqref{newrealkeyrate}, we must consider now how
to use the decoy state idea to estimate $Q_0$, $Q_1$, $e_1$, $Q_2$
and $e_2$. A similar problem for orthogonal encoding protocols has
been analyzed explicitly by Lo and his co-workers in \cite{mqzl}.
Here we exploit their method to solve the question in
nonorthogonal protocols.

For simplicity, we propose a specific protocol that uses only four
decoy states: vacuum and three weak decoy states. The vacuum can
be used to estimate the background rate,
\begin{eqnarray} \label{y0e0}
Y_0=Q_{vacuum}, \nonumber\\
e_0=E_{vacuum}=\frac{1}{2}.
\end{eqnarray}
The dark counts occur randomly; thus the error rate of the the
dark count is $1/2$. The signal and three decoy states with
expected numbers $\mu$, $\nu_1$, $\nu_2$ and $\nu_3$ satisfy
\begin{eqnarray} \label{munv}
0<\nu_3<\nu_2\leq \frac{2}{3}\mu<\nu_1\leq \frac{3}{4} \mu, \nonumber\\
\nu_1+\nu_2>\mu,\nonumber\\
\nu_2+\nu_3<\mu.
\end{eqnarray}
Alice and Bob will get the following gains and QBERs for signal
state and these three decoy states:
\begin{equation}\label{Decoy:SigDecoym}
\begin{aligned}
Q_{\mu}e^{\mu} &= Y_0+Y_1\mu+\frac{Y_2\mu^2}{2}+\sum_{i=3}^{\infty} Y_i\frac{\mu^i}{i!}, \\
E_{\mu}Q_{\mu}e^{\mu} &= e_0Y_0+e_1Y_1\mu+\frac{e_2Y_2\mu^2}{2}+\sum_{i=3}^{\infty} e_iY_i\frac{\mu^i}{i!}, \\
Q_{\nu_1}e^{\nu_1} &= Y_0+Y_1\nu_1+\frac{Y_2\nu_1^2}{2}+\sum_{i=3}^{\infty}Y_i\frac{\nu_1^i}{i!},  \\
E_{\nu_1} Q_{\nu_1} e^{\nu_1} &= e_0Y_0+e_1Y_1\nu_1+\frac{e_2Y_2\nu_1^2}{2}+\sum_{i=3}^{\infty}e_iY_i\frac{\nu_1^i}{i!}, \\
Q_{\nu_2}e^{\nu_2} &= Y_0+Y_1\nu_2+\frac{Y_2\nu_2^2}{2}+\sum_{i=3}^{\infty}Y_i\frac{\nu_2^i}{i!},  \\
E_{\nu_2} Q_{\nu_2} e^{\nu_2} &= e_0Y_0+e_1Y_1\nu_2+\frac{e_2Y_2\nu_2^2}{2}+\sum_{i=3}^{\infty}e_iY_i\frac{\nu_2^i}{i!}, \\
Q_{\nu_3}e^{\nu_3} &= Y_0+Y_1\nu_3+\frac{Y_2\nu_3^2}{2}+\sum_{i=3}^{\infty}Y_i\frac{\nu_3^i}{i!},  \\
E_{\nu_3} Q_{\nu_3} e^{\nu_3} &=
e_0Y_0+e_1Y_1\nu_3+\frac{e_2Y_2\nu_3^2}{2}+\sum_{i=3}^{\infty}e_iY_i\frac{\nu_3^i}{i!}.
\end{aligned}
\end{equation}

Alice and Bob can estimate the lower bound of $Y_1$ and the upper
bound of $e_1$ from Eq. \eqref{Decoy:SigDecoym} by using  decoy
states $\nu_2$ and $\nu_3$ . The lower bound of $Y_1$ is given by
\begin{eqnarray} \label{y1}
&Q_{\nu_2}e^{\nu_2}-Q_{\nu_3}e^{\nu_3}= Y_1(\nu_2-\nu_3)+\sum_{i\geq 2}\frac{Y_i}{i!}(\nu_2^i-\nu_3^i)\nonumber\\
&\leq Y_1(\nu_2-\nu_3)+\frac{\nu_2^2-\nu_3^2}{\mu^2}\sum_{i\geq 2}\frac{Y_i\mu^i}{i!}\nonumber\\
&=Y_1(\nu_2-\nu_3)+\frac{\nu_2^2-\nu_3^2}{\mu^2}(Q_{\mu}e^{\mu}-Y_0-Y_1\mu).
\end{eqnarray}
Here,in order to prove the inequality in Eq. \eqref{y1}, we have
made use of the inequality that $a^i-b^i\leq a^2-b^2$ whenever
$0<b<a\leq \frac{2}{3}$, and $i\geq 2$. The last equality sign
holds in the in Eq. \eqref{y1} if and only if Eve raises the yield
of two-photon states and blocks all the states with photon number
greater than $2$. In fact Eve will not take this tactics because
she cannot achieve full information on two-photon state. The upper
bound of $e_1$ is given by
\begin{eqnarray} \label{e1}
&E_{\nu_3} Q_{\nu_3} e^{\nu_3} = e_0Y_0+e_1Y_1\nu_3+\sum_{i=2}^{\infty}e_iY_i\frac{\nu_3^i}{i!} \nonumber\\
&\geq e_0Y_0+e_1Y_1\nu_3.
\end{eqnarray}
By solving Eq. \eqref{y1} and Eq. \eqref{e1}, the lower bound of
$Y_1$ and upper bound of $e_1$ are given by
\begin{eqnarray} \label{lowery1}
Y_1 &\geq &  Y_1^{L}
\nonumber\\
&=&\frac{\mu^2(Q_{\nu_2}e^{\nu_2}-Q_{\nu_3}e^{\nu_3})-(\nu_2^2-\nu_3^2)(Q_{\mu}e^{\mu}-Y_0)}{\mu^2(\nu_2-\nu_3)(\mu-\nu_2-\nu_3)},\nonumber\\
e_1&\leq & e_1^U=\frac{E_{\nu_3} Q_{\nu_3}
e^{\nu_3}-e_0Y_0}{Y_1^{L}\nu_3}.
\end{eqnarray}
Then, according $Q_1=Y_1P_1(\mu)$, the gain of single-photon
states is given by
\begin{eqnarray} \label{lowerq1}
Q_1&\geq & Q_1^L=Y_1^L\mu e^{-\mu}.
\end{eqnarray}

Next, Alice and Bob can estimate the lower bounds of $Y_2$ and the
upper bound of $e_2$ respectively by using  decoy states $\nu_1$,
$\nu_2$ and $\nu_3$ from Eq. \eqref{Decoy:SigDecoym} under
conditions Eq. \eqref{munv}. The lower bound of $Y_2$ is given by
\begin{eqnarray} \label{y2}
&Q_{\nu_1}e^{\nu_1}-Q_{\nu_2}e^{\nu_2}\nonumber\\
&= Y_1(\nu_1-\nu_2)+\frac{Y_2}{2}(\nu_1^2-\nu_2^2)+\sum_{i\geq 3}\frac{Y_i}{i!}(\nu_1^i-\nu_2^i)\nonumber\\
&\leq Y_1(\nu_1-\nu_2)+\frac{Y_2}{2}(\nu_1^2-\nu_2^2)+\frac{\nu_1^3-\nu_2^3}{\mu^3}\sum_{i\geq 3}\frac{Y_i\mu^i}{i!}\nonumber\\
&=Y_1(\nu_1-\nu_2)+\frac{Y_2}{2}(\nu_1^2-\nu_2^2)\nonumber\\
&+\frac{\nu_1^3-\nu_2^3}{\mu^3}\sum_{i\geq 3}(Q_{\mu}e^{\mu}-Y_0-Y_1\mu-\frac{Y_2\mu^2}{2})\nonumber\\
&=\frac{Y_2}{2}(\nu_1^2-\nu_2^2)+\frac{\nu_1^3-\nu_2^3}{\mu^3}\sum_{i\geq
3}(Q_{\mu}e^{\mu}-Y_0-\frac{Y_2\mu^2}{2}).
\end{eqnarray}
In order to prove the inequality in Eq. \eqref{y2}, we have made
use of the inequality that $a^i-b^i\leq a^2-b^2$ whenever
$0<b<a\leq \frac{3}{4}$, and $i\geq 3$. The last equality sign
holds in Eq. \eqref{y2} if and only if Eve raises the yield of
three-photon states and blocks all the states with photon number
greater than $3$. In addition, to obtain the last sign equality in
Eq. \eqref{y2}, we have let $\nu_1$ and $\nu_2$ satisfying
\begin{equation}\label{nu12}
\nu_1-\nu_2-\frac{\nu_1^3-\nu_1^3}{\mu^2}=0.
\end{equation}
The upper bound of $e_2$ is given by
\begin{eqnarray} \label{e2}
&E_{\nu_3} Q_{\nu_3} e^{\nu_3} = e_0Y_0+e_1Y_1\nu_3+\frac{e_2Y_2\nu_3^2}{2}+\sum_{i=3}^{\infty}e_iY_i\frac{\nu_3^i}{i!} \nonumber\\
&\geq e_0Y_0+\frac{e_2Y_2\nu_3^2}{2}.
\end{eqnarray}
By solving Eq. \eqref{y2} and Eq. \eqref{e2}, the lower bound of
$Y_2$ and $Q_2$ and upper bound of $e_2$ are given by
\begin{eqnarray} \label{lowery2}
Y_2 &\geq &  Y_2^{L}
\nonumber\\
&=&\frac{2\mu(Q_{\nu_1}e^{\nu_1}-Q_{\nu_2}e^{\nu_2})-2(\nu_1-\nu_2)(Q_{\mu}e^{\mu}-Y_0)}{\mu(\nu_1-\nu_2)(\nu_1+\nu_2-\mu)},\nonumber\\
Q_2 &\geq & Q_2^{L}=\frac{Y_2^{L}\mu^2 e^{-\mu}}{2} ,\nonumber\\
e_2&\leq & e_2^U=\frac{2E_{\nu_3} Q_{\nu_3}
e^{\nu_3}-2e_0Y_0}{Y_2^{L}\nu_3^2}.
\end{eqnarray}

Now, the lower bound of the secure key generation rate, according
to Eq. \eqref{newrealkeyrate}, is given by:
\begin{eqnarray}\label{lowerRate}
 S^L = q  \{-Q_{\mu}f(E_{\mu})H_2(E_{\mu})+Q_0+Q_1^L[ 1- H_2(e_1^U)]\nonumber \\
 +Q_2^L[ 1-H_2(e_2^U)]\},
\end{eqnarray}
where $Q_0=Y_0e^{-\mu}=Q_{vacuum}e^{-\mu}$. Comparing our result
(given in Eq. \eqref{newrealkeyrate}) with the prior result in
\cite{mqzl}(given in Eq. \eqref{practicalkeyrate}), we see that
the main difference is that in our result, two additional terms,
$Q_0$ and $Q_2^L[ 1-H_2(e_2^U)]$, can also generate secure keys.
To fix the ideas, we will compare our protocol with the SARG04
protocol and BB84 protocol according Eqs. \eqref{newrealkeyrate}
\eqref{worst} and \eqref{practicalkeyrate} respectively in the
following paragraph.

\begin{figure}
\centering \resizebox{8cm}{!}{\includegraphics{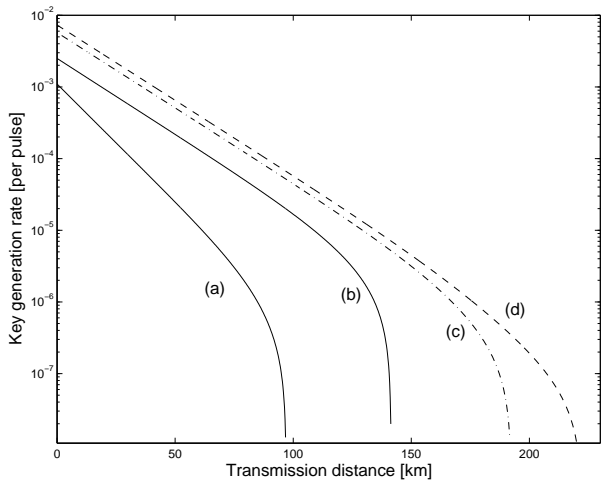}}
\caption{(a)The optimal secure generation rate for SARG04 protocol
without decoy states, (b)The optimal secure generation rate for
BB84 protocol with decoy states ($\mu=0.48$), (c) The secure
generation rate for our protocol by using the formula
\eqref{lowerRate}($\mu=0.48$), (d) The secure generation rate for
our protocol by using the formula \eqref{lowerRate} ($\mu=0.30$).
The parameters is given according to experiment GYS \cite{GYS}:
$\alpha=0.21dB/km$, $e_{det}=3.3\%$, $Y_0=1.7\times 10^{-6}$, and
the detection efficiency of Bob's setup $\eta_Bob=0.045$.
$f(E_{\mu})=1.22$.} \label{fig1}
\end{figure}

For simplicity, We only consider the asymptotic case (i.e. omit
statical fluctuations of $Q_n$ and $e_n$). By using the GYS
\cite{GYS} experiment as an example, the result shows in Fig.~1.
The curve (a) is the optimal secure generation rate for SARG04
protocol without decoy states achieved by using Eq. \eqref{worst}.
The curve (b) is a simple repeat of Ref. \cite{lmc} for BB84
protocol with decoy states. We note that our protocol is better
than both SARG04 protocol without decoy states and Lo's protocol
at any distance. The maximal distances of the three protocols are
220, 142, and 97 km respectively. Theoretically, we can achieve a
longer transmission distance with our method when we decrease the
value of $\mu$. In these cases, however, the weak decoy state
method cannot work efficiently due to the statical fluctuations.

In summary, we have proposed an efficient and feasible
nonorthogonal decoy-state protocol to do QKD over very lossy
channel. we have clearly demonstrated how to estimate the lower
bound of the secure key generation rate in this new protocol. Our
result shows that, the combination of decoy state method and
nonorthogonal states encoding protocol can make great progress at
the secure key generation rate. Our protocol can be realized
easily because it is the same as Lo's protocol in operation.

J.-B.Li thanks Xiongfeng Ma for his kind help with numerical
calculations. This work is supported by Scientific Research Fund
of Hunan Provincial Education Department No. 03c213.


\end{document}